\documentclass[%
 reprint,
 amsmath,amssymb,
 aps,
 pra,
 superscriptaddress,
 floatfix,
]{revtex4-2}

\usepackage{algorithm}
\usepackage{algpseudocode}
\usepackage{array}
\usepackage{bm}
\usepackage[justification=raggedright,singlelinecheck=false]{caption}
\usepackage{graphicx}
\usepackage{makecell}
\usepackage{multirow}
\usepackage{physics}
\usepackage{braket}
\usepackage{txfonts}
\usepackage{url}
\usepackage{xcolor}
\newcommand{\upom}{UPOM}
\newcommand{\duupom}{DU-UPOM}
\newcommand{\lom}{LOM}
\newcommand{\R}{\mathbb{R}}
\newcommand{\MSE}{\mathrm{MSE}}

\begin{document}

\preprint{APS/123-QED}

\title{Slack-Free Deep-Unfolded Combinatorial Optimization Solver for Inequality Constraints}
\author{Ryo Hagiwara}
\affiliation{Institute of Science Tokyo, Ookayama, Tokyo 152-8550, Japan}
\author{Shunta Arai}
\affiliation{Institute of Science Tokyo, Ookayama, Tokyo 152-8550, Japan}
\author{Satoshi Takabe}
\affiliation{Institute of Science Tokyo, Ookayama, Tokyo 152-8550, Japan}
\affiliation{RIKEN Center for Advanced Intelligence Project (AIP), Nihonbashi, Tokyo 103-0027, Japan}
\date{\today}

  \begin{abstract}

      {Quantum annealing (QA) is used to solve combinatorial optimization problems (COPs).
  When COPs are implemented on quantum annealers, they are typically encoded as quadratic unconstrained binary optimization (QUBO) problems, but constraint encodings often increase the number of qubits and the embedding overhead.}
  This issue is particularly important for COPs with inequality constraints, where standard slack-variable formulations introduce additional binary variables.
  Unbalanced penalization (UP) avoids slack variables, but the original UP formulation requires tuning two penalty coefficients and contains a squared residual term that can increase the number of quadratic
  couplings. In this paper, we propose the {u}nbalanced {p}enalization Ohzeki method (UPOM), which combines UP with the Ohzeki method for inequality-constrained COPs.
  UPOM replaces the two static penalty coefficients of original UP with an auxiliary-variable update and removes the squared residual term from {the Hamiltonian used for sampling}.
  We further propose the deep-unfolded {u}nbalanced {p}enalization Ohzeki method (DU-UPOM), which learns the step-size schedule of the UPOM update from training instances.
  Numerical experiments on random knapsack {problems} show that UPOM improves over original UP and that DU-UPOM reaches optimal solutions in fewer iterations than fixed-step UPOM and {other baseline.}
  These results demonstrate that the proposed framework reduces the tuning and embedding burdens of UP while making the Ohzeki method trainable for inequality constraints.

  \end{abstract}

\maketitle

\section{\label{sec:introduction}Introduction}

Combinatorial optimization problems (COPs) ask for the minimum or maximum of an objective function over discrete variables subject to constraints.
Typical examples are the traveling salesman problem and the knapsack problem.
Because COPs are closely related to the P versus NP problem, obtaining an exact optimum in polynomial time is difficult for many problem classes~\cite{PNP}.

Representative nondeterministic approaches to such problems include simulated annealing (SA)~\cite{SA} and quantum annealing (QA)~\cite{QA}.
SA searches for low-energy configurations using {Markov-chain} {Monte-Carlo} (MCMC) sampling with thermal fluctuations~\cite{Metropolis,Hastings}.
QA instead uses quantum fluctuations and has attracted research attention as both a sampler and a COP solver, including comparisons with SA~\cite{QAvsSA}.
QA has been implemented in specialized devices called quantum annealers, such as D-Wave systems~\cite{DwaveSys}.

To run a COP on a quantum annealer, the problem is usually encoded as a quadratic unconstrained binary optimization (QUBO) problem.
This encoding must satisfy hardware restrictions such as a finite number of qubits and sparse fixed connectivity.
When constraints are represented by quadratic penalty terms, the resulting QUBO can contain many interactions.
Minor embedding of these interactions onto the hardware graph may require many physical qubits for one logical variable, which limits the size of problem instances that can be executed on a quantum annealer~\cite{MinerEmbedding1,MinerEmbedding2,MinerEmbedding3,MinerEmbedding4}.

To reduce the overhead caused by such penalty terms, the Ohzeki method was proposed for linearly constrained COPs~\cite{Ohzeki}.
Using the {Hubbard--Stratonovich} transformation~\cite{Hubbard,Stratonovich}, the method introduces auxiliary variables and avoids placing the squared penalty term directly in {the Hamiltonian used for sampling}.
At each iteration, a sampler estimates the expectation values of constraint functions, and the auxiliary variables are updated so that these expectations approach the constraint values.

  The method is especially suitable for QA because it can reduce the number of interactions that must be embedded, while the same framework can in principle be combined with any sampler, including MCMC, SA,
  and simulated quantum annealing~\cite{SQA,SQA2}.
  In practical implementations, the convergence speed and solution quality depend on the sampler, inverse temperature, and update parameters.
  In particular, the step sizes in the auxiliary-variable update are important tunable parameters.

% However, the convergence speed and solution quality depend strongly on the step sizes used in the auxiliary variable update.

Deep unfolding (DU) is a framework that converts an iterative algorithm into a trainable layered model~\cite{DU,DU2}.
Each iteration is interpreted as one layer, and algorithmic parameters such as step sizes are made layer dependent and optimized from {a dataset of} problem instances.
Unlike a neural network, DU keeps the update rule of the original algorithm and learns only a small number of internal parameters that control convergence.

DU has been used in a wide range of model-based learning problems, including signal processing~\cite{Monga,Boyd}, wireless communications and signal detection~\cite{DUCS,THS}, compressed sensing~\cite{TISTA},
image reconstruction~\cite{DUIM}, and algorithms inspired by quantum computation~\cite{TSB,DULQA}.

% DU has been used in a wide range of learning problems based on models, including signal processing~\cite{Monga,Boyd}, wireless communications~\cite{DUCS}, compressed sensing~\cite{TISTA}, image reconstruction and super resolution~\cite{DUIM}, signal detection~\cite{THS}, and algorithms inspired by quantum computation~\cite{TSB,DULQA}.
% This broad applicability makes DU a natural approach for learning the step size schedule of the Ohzeki method.

In our previous study, we addressed the step size {scheduling} problem by combining the Ohzeki method with DU, and proposed the deep unfolded Ohzeki method (DUOM)~\cite{duom}.
DUOM regards each iteration of the Ohzeki method as a layer and learns the step size sequence from a dataset of problem instances.
Although MCMC samplers {involve} nondifferentiable {acceptance decisions}, the derivatives {required for backpropagation can be estimated} from sampling moments.
The numerical results in Ref.~\cite{duom} showed that learning the step sizes accelerates the convergence of the Ohzeki method with a classical sampler.

We previously adapted this learning-based solver to QA through {classical-quantum} transfer learning~\cite{cqtl}.
Specifically, the step sizes of DUOM were trained using a classical sampler and then transferred to the Ohzeki method implemented on a quantum annealer.
The experiments showed that the transferred step sizes improve {the} convergence speed and approximation performance in QA execution.
These two studies established a trainable COP solver based on the Ohzeki method and demonstrated its usefulness for QA.

A remaining limitation is that these previous studies based on DUOM have not yet been applied to problems with inequality constraints.
Many COPs, including the binary knapsack problem~\cite{MartelloToth1990} studied in this paper, are naturally formulated with inequality constraints.
The difficulty comes from the fact that the previous DUOM formulation was constructed for equality constraints, where each constraint function is driven toward a prescribed value.
Therefore, extending the Ohzeki {method} based on learning to inequalities is an important step toward a broader trainable solver for COPs.

A standard way to handle an inequality in QUBO form is to introduce slack variables and convert the inequality into an equality~\cite{KomiyamaSuzuki2024}.
The Ohzeki method has also been applied to problems with inequality constraints through such formulations with slack variables~\cite{YuNabil2021}.

  However, slack variables must themselves be represented by additional binary variables.
  This increases the number of logical variables and, after minor embedding, can increase the number of physical qubits required on a quantum annealer.
  Thus, although slack variables allow inequalities to be treated as equalities, they introduce an additional qubit overhead and can offset part of the hardware advantage expected from removing quadratic penalty
  terms in the Ohzeki method.

Recently, unbalanced penalization (UP) was proposed as an approach to inequality constraints without slack variables~\cite{MontanezBarrera2024}.
Original UP handles an inequality by adding an asymmetric penalty with two coefficients, $\lambda_1$ and $\lambda_2$, and includes a squared residual term.
Because no slack variables are introduced, UP is attractive from the viewpoint of qubit usage.
At the same time, two issues remain.
First, the two hyperparameters, $\lambda_1$ and $\lambda_2$, must be tuned jointly, which is difficult because {suitable values depend on the characteristics of the target problem and the sampler}.
Second, the squared residual term produces additional quadratic interactions, which can again increase the embedding burden and the number of physical qubits needed on a quantum annealer.

In this paper, we combine UP with the Ohzeki method and propose the {unbalanced penalization} Ohzeki method (\upom).
The proposed method keeps the UP formulation without slack variables, but replaces the two independent static UP hyperparameters with an auxiliary variable update of the Ohzeki method governed by a step size schedule.
Moreover, {the Hamiltonian used for sampling in \upom} does not contain the squared residual term.
Thus, \upom\ reduces the tuning difficulty of original UP and removes the corresponding quadratic interactions from the sampler while preserving the advantage of using no slack variables.

We further combine \upom\ with DU and propose the deep unfolded {unbalanced penalization} Ohzeki method (\duupom).
In \duupom, the step-size schedule {for} the auxiliary-variable update in \upom\ is learned automatically from training instances.
Together with our previous studies on DUOM and {classical-quantum} transfer learning~\cite{duom,cqtl}, \duupom\ extends the trainable Ohzeki {method} from equality-constrained COPs to inequality-constrained COPs.

The contributions of this work are threefold.
First, we improve original UP by replacing its two static hyperparameters with an auxiliary variable update of the Ohzeki method and by removing the squared residual term from {the Hamiltonian used for sampling}.
Second, we formulate an application of the Ohzeki method to COPs with inequality constraints without using slack variables.
Third, we apply DU to this Ohzeki {method} for inequality constraints and evaluate the resulting \duupom\ on random knapsack instances.
The experiments show that \upom\ avoids the two-parameter search required by original UP and that \duupom\ substantially improves the convergence speed of \upom.

The remainder of this paper is organized as follows.
Sec.~\ref{sec:background} reviews the Ohzeki method, DU, DUOM, and original UP.
Sec.~\ref{sec:upom} formulates \upom\ by combining UP with the Ohzeki method and compares it numerically with original UP.
Sec.~\ref{sec:duupom} applies DU to \upom, describes the learning and evaluation settings, and compares \duupom\ with \upom\ using a fixed step size, \lom, and the Ohzeki method with slack variables.
Sec.~\ref{sec:conclusion} concludes the paper and discusses future directions.

\section{\label{sec:background}Background and Related Work}

\subsection{\label{subsec:ohzeki}Ohzeki method}

The Ohzeki method is a solver based on sampling for binary optimization problems with linear equality constraints.
We consider a binary optimization problem
\begin{equation}
  \begin{aligned}
  \min_{\bm{x}\in\{0,1\}^{N}} \quad & f_0(\bm{x}) \\
  \textrm{s.t.} \quad & f_k(\bm{x})=C_k,
  \quad k=1,\ldots,m ,
  \end{aligned}
  \label{eq:equality_cop}
\end{equation}
where $N$ denotes the number of binary decision variables, $m$ denotes the number of constraints, $f_0$ is the objective function, $f_k$ is the linear function defining the
  $k$-th constraint, and $C_k$ is its prescribed constraint value.
  A direct penalty formulation uses
\begin{equation}
  L_{\lambda}(\bm{x})
  =
  f_0(\bm{x})
  +\lambda\sum_{k=1}^{m}\left(f_k(\bm{x})-C_k\right)^2 ,
  \label{eq:penalty_form}
\end{equation}
where $\lambda>0$ is a penalty coefficient that controls the strength of the constraint penalty.
For linear $f_k$, the squared term generally creates pairwise couplings among many variables.

 By applying the Hubbard--Stratonovich transformation to the quadratic penalty term, the Ohzeki method introduces auxiliary variables and samples from
\begin{equation}
  Q(\bm{x};\bm{v}^{(t)})
  =
  \frac{1}{Z(\bm{v}^{(t)})}
  \exp\left[
    -\beta f_0(\bm{x})
    +\beta\sum_{k=1}^{m}v_k^{(t)} f_k(\bm{x})
  \right],
  \label{eq:ohzeki_distribution}
\end{equation}
where $\bm{v}^{(t)}\in\R^m$ are auxiliary variables and $\beta$ is the inverse temperature.
The auxiliary variables are updated by
\begin{equation}
  v_k^{(t+1)}
  =
  v_k^{(t)}
  +\eta_t
  \left(
    C_k-\Braket{f_k(\bm{x})}_{Q(\bm{x};\bm{v}^{(t)})}
  \right),
  \label{eq:ohzeki_update}
\end{equation}
where $\eta_t$ is the step size and $\Braket{f_k(\bm{x})}_{Q(\bm{x};\bm{v}^{(t)})}$ denotes the expectation value of $f_k(\bm{x})$ with respect to the distribution $Q(\bm{x};
  \bm{v}^{(t)})$.
  The sampling and auxiliary-variable update are repeated for a preset number of iterations.
  After the final iteration, candidate solutions are sampled from the final distribution and evaluated with the original constrained objective.
  Algorithm~\ref{alg_oh} summarizes this procedure.

\begin{figure}[!t]
\begin{algorithm}[H]
    \captionsetup{justification=raggedright, singlelinecheck=false}
  \caption{Ohzeki method}\label{alg_oh}
  \begin{algorithmic}[1]
  \State \textbf{Input:} $\beta$,$\lambda$,$f_0$, $\{f_k, C_k\}$, max iteration: $T$
  \State \textbf{Initialize:} $\bm{v}^{(0)} \in \mathbb{R}^m$
  \For{ $t = 0, 1, \dots ,T-1$}
  \State estimate $\{\braket{f_k(\bm{x})}_{Q(\bm{x}; \bm{v}^{(t)})}\}$ by sampling from
  \State \quad $Q(\bm{x}; \bm{v}^{(t)}) \propto \exp(-\beta f_0(\bm{x}) + \beta \sum_{k=1}^{m} v_{k}^{(t)} f_k(\bm{x}))$
  \State update $\bm{v}^{(t+1)}$ by
  \State \quad $v_k^{(t+1)} = v_k^{(t)} + \eta_{t} ( C_k - \braket{f_k(\bm{x})}_{Q(\bm{x}; \bm{v}^{(t)})})$
  \EndFor\label{euclidendwhile}
  \State \textbf{return}  $\arg\min L(\bm{x}; \lambda)$ by sampling from $Q(\bm{x}; \bm{v}^{(T)})$
  \end{algorithmic}
\end{algorithm}
 \end{figure}

\subsection{\label{subsec:deep_unfolding}Deep unfolding}

Deep unfolding is a learning framework for improving an iterative algorithm by embedding trainable parameters into its update rule and optimizing them by backpropagation~\cite{DU,DU2}.
The typical setting is that many random instances from the same problem class must be solved repeatedly, and the available dataset represents the statistical structure of that class.
This assumption is natural in signal processing and is also useful for families of combinatorial optimization instances.
Under this setting, DU preserves the form of the original algorithm while learning internal parameters that strongly affect convergence.

As a simple example, {we} consider gradient descent for minimizing a differentiable function $g(\bm{z}^{(t)})$, where $\bm{z}^{(t)}$ denotes the variable at iteration $t$.
Starting from an initial value $\bm{z}^{(0)}$, the standard update is $\bm{z}^{(t+1)}=\bm{z}^{(t)}-\eta\nabla g(\bm{z}^{(t)})$ for $t=0,\ldots,T-1$, where $\eta$ is the step size
  and $T$ is the number of iterations.

  The choice of $\eta$ has a strong influence on the convergence behavior of the algorithm.
  {If $\eta$ is too large, convergence can become unstable; if it is too small, convergence can be slow.}

% DU unfolds the $T$ iterations into a layered model and replaces the fixed step size by parameters that depend on the layer,
DU then regards the $T$ iterations of the update rule as $T$ layers.
  Instead of using a fixed step size for all iterations, {DU allows the step size to vary across layers as follows:}
\begin{equation}
  \bm{z}^{(t+1)}
  =
  \bm{z}^{(t)}-\eta_t \nabla g(\bm{z}^{(t)}).
  \label{eq:dugd}
  \nonumber
\end{equation}

In this example, the trainable parameters are the step-size sequence $\{\eta_t\}_{t=0}^{T-1}$.
The unfolded model keeps the computational structure of the original gradient descent algorithm, while only the step-size schedule is optimized.
{Thus, the learned parameters directly correspond to the step sizes in the original algorithm and are therefore easy to interpret, while the number of trainable parameters is limited to $T$.}

In a supervised DU setting, a dataset
$\{\bm{z}_0^{(d)},g^{(d)},\bm{z}_\ast^{(d)}\}_{d=1}^{D}$
is available.
Here $g^{(d)}$ is the objective of the $d$-th instance, $\bm{z}_0^{(d)}$ is the initial point, and $\bm{z}_\ast^{(d)}$ is an optimal solution.
After applying Eq.~\eqref{eq:dugd} for $T$ layers, the output is denoted by $\bm{z}_T^{(d)}$.
A typical training loss is the mean squared error
\begin{equation}
  L_{\MSE}
  =
  \frac{1}{D}
  \sum_{d=1}^{D}
  \left\|
    \bm{z}_T^{(d)}-\bm{z}_\ast^{(d)}
  \right\|_2^2 .
  \label{eq:du_mse}
\end{equation}
The gradients $\partial L_{\MSE}/\partial \eta_t$ are then computed by backpropagation through the unfolded iterations, {and parameters $\eta_t$ are updated by an optimizer.}

    As a result, DU learns a step-size schedule adapted to the instance distribution and tends to improve convergence compared with the original algorithm~\cite{Chebyshev}.

\subsection{\label{subsec:duom_background}Deep unfolded Ohzeki method}

DUOM applies the DU principle to the Ohzeki method~\cite{duom}.
{Since the step size $\eta_t$ in Eq.~\eqref{eq:ohzeki_update} controls the auxiliary-variable update and strongly affects convergence, DUOM treats $\Theta=\{\eta_t\}_{t=0}^{T-1}$ as the set of trainable parameters.}
Unlike the supervised example in Sec.~\ref{subsec:deep_unfolding}, DUOM is trained without requiring optimal solutions in advance.
This is important for constrained COPs because obtaining exact optima for all training instances is generally expensive.
{For notational simplicity, we abbreviate $Q(\bm{x};\bm{v}^{(t)})$ as $Q$, with the iteration determined by context.}
% Instead, DUOM minimizes the expected penalized objective after the final Ohzeki layer,
Instead, DUOM minimizes the expected penalized objective with respect to the sampling distribution obtained after $T$ iterations,
\begin{equation}
  L_{\rm DUOM}
  =
  \Braket{
    L_{\lambda}(\bm{x})
  }_{{Q}}.
  \label{eq:duom_loss}
\end{equation}
The step sizes are updated by a standard optimizer such as Adam~\cite{Adam}.

The main technical point is that the sampling step is not differentiable.
For example, {Metropolis--Hastings~(MH)} sampling uses discrete {acceptance} decisions~\cite{Metropolis,Hastings}, so {standard} automatic differentiation cannot directly pass through the sampler.
The gradients for the trainable step sizes are therefore evaluated by backpropagation through the unfolded update equations.
For the update in Eq.~\eqref{eq:ohzeki_update}, the chain rule gives
\begin{align}
  \frac{\partial L_{\rm DUOM}}{\partial \eta_t}
  &=
  \sum_{k=1}^{m}
  \frac{\partial L_{\rm DUOM}}{\partial v_k^{(T)}}
  \left(
    \prod_{u=t+1}^{T-1}
    \frac{\partial v_k^{(u+1)}}{\partial v_k^{(u)}}
  \right)
  \frac{\partial v_k^{(t+1)}}{\partial \eta_t},
  \label{eq:duom_backprop_eta}\\
  \frac{\partial v_k^{(u+1)}}{\partial v_k^{(u)}}
  &=
  1-\eta_u
  \frac{\partial
  \Braket{f_k(\bm{x})}_{{Q}}}
  {\partial v_k^{(u)}} .
  \label{eq:duom_backprop_layer}
\end{align}
The remaining factor is
\begin{equation}
  \frac{\partial v_k^{(t+1)}}{\partial \eta_t}
  =
  C_k-\Braket{f_k(\bm{x})}_{{Q}}.
  \label{eq:duom_backprop_eta_factor}
\end{equation}
{Equations~\eqref{eq:duom_backprop_eta}-\eqref{eq:duom_backprop_eta_factor}} are the backpropagation equations used in DUOM.
However, Eq.~\eqref{eq:duom_backprop_layer} still contains the derivative of an expectation estimated by a nondifferentiable sampler.
DUOM avoids this difficulty by estimating this derivative from the same samples used in the forward pass.
For the distribution in Eq.~\eqref{eq:ohzeki_distribution},
\begin{equation}
  \frac{\partial}{\partial v_l}
  \Braket{f_k(\bm{x})}_{Q}
  =
  \beta\left(
    \Braket{f_k(\bm{x})f_l(\bm{x})}_{Q}
    -
    \Braket{f_k(\bm{x})}_{Q}
    \Braket{f_l(\bm{x})}_{Q}
  \right).
  \label{eq:moment_derivative}
\end{equation}
Thus, the Jacobian of the expectation is given by the covariance matrix of the constraint functions.
For the single constraint, which is the setting used in the numerical experiments in this paper, Eq.~\eqref{eq:moment_derivative} takes the simpler form
\begin{equation}
  \frac{\partial}{\partial v}
  \Braket{f(\bm{x})}_{Q}
  =
  \beta
  \operatorname{Var}_{Q}\!\left[f(\bm{x})\right].
  \label{eq:duom_variance}
\end{equation}
Substituting this variance estimate obtained from sampling into Eq.~\eqref{eq:duom_backprop_layer} enables backpropagation without automatic differentiation through the sampler.
Using this variance expression, the derivative of one step of the auxiliary-variable recursion in Eq.~\eqref{eq:ohzeki_update} is given by
\begin{equation}
  \frac{\partial v^{(t+1)}}{\partial v^{(t)}}
  =
  1-\eta_t\beta
  \operatorname{Var}_{{Q}}
  \!\left[f(\bm{x})\right].
  \label{eq:duom_layer_derivative}
\end{equation}
Combining these derivatives by the chain rule yields the gradient of Eq.~\eqref{eq:duom_loss} with respect to each $\eta_t$.
Schematic diagrams and further implementation details of DUOM and its extension by classical to quantum transfer learning are given in Refs.~\cite{duom,cqtl}.
% In Ref.~\cite{duom}, the effectiveness of DUOM learning was demonstrated for the problem of selecting the $K$ minimum set from $N$ random variables and for image reconstruction problems solved with an MCMC sampler.
% In Ref.~\cite{cqtl}, classical to quantum transfer learning was introduced, in which step sizes trained with a classical sampler are transferred to the Ohzeki method executed with a quantum annealer.
% The results showed that the learned DUOM step sizes also accelerate convergence when the Ohzeki method is executed on a quantum annealer.

Previous DUOM studies were formulated for equality constraints.
The present work uses the same differentiability mechanism based on moments, but moves the target problem to inequality constraints by combining UP with the Ohzeki method.
The key requirement is that the inequality formulation retain a sampler distribution whose expectation derivatives are available from variances and covariances, as in Eq.~\eqref{eq:moment_derivative}.

\subsection{\label{subsec:up_background}Unbalanced penalization}

Original UP was introduced for binary optimization problems with inequality constraints.
For a single inequality, we consider
\begin{equation}
  \begin{aligned}
  \min_{\bm{x}\in\{0,1\}^{N}} \quad & f_0(\bm{x}) \\
  \textrm{s.t.} \quad & F(\bm{x}) \le B ,
  \end{aligned}
  \label{eq:ineq_constraint}
\end{equation}
where $F(\bm{x})$ is a linear constraint function, and $B$ is a prescribed upper bound of the constraint.
The residual of the inequality is defined as
\begin{equation}
  h(\bm{x})=B-F(\bm{x}).
  \label{eq:residual}
\end{equation}
The feasible region is $h(\bm{x})\ge0$.
Original UP encodes the inequality by the penalty
\begin{equation}
  \zeta_{\rm UP}(\bm{x})
  =
  -\lambda_1 h(\bm{x})
  +\lambda_2 h(\bm{x})^2,
  \quad
  \lambda_1,\lambda_2>0 .
  \label{eq:up_penalty}
\end{equation}
This form can be interpreted as a second order approximation of an asymmetric exponential penalty~\cite{MontanezBarrera2024}.
The linear term makes configurations with large positive residual favorable, whereas the quadratic term penalizes violations.
Original UP then uses the Hamiltonian
\begin{equation}
  H_{\rm UP}(\bm{x})
  =
  f_0(\bm{x})
  -\lambda_1\left(B-F(\bm{x})\right)
  +\lambda_2\left(B-F(\bm{x})\right)^2 .
  \label{eq:up_hamiltonian}
\end{equation}

The main advantage of Eq.~\eqref{eq:up_hamiltonian} is that no slack variable is introduced.
This contrasts with the standard encoding using slack variables, which rewrites $F(\bm{x})\le B$ as $F(\bm{x})+S=B$ and then represents $S$ by additional binary variables~\cite{KomiyamaSuzuki2024}.
If binary encoding is used, at least $\lceil\log_2(L+1)\rceil$ slack bits are needed to represent a slack range $0\le S\le L$.
The number of logical variables and, after embedding, the number of physical qubits are therefore increased.

Original UP avoids this variable increase, but it introduces two static hyperparameters, $\lambda_1$ and $\lambda_2$.
Their joint tuning is difficult because {suitable values} depend on {the characteristics of the target problem} and on the sampler.
Moreover, the squared term in Eq.~\eqref{eq:up_hamiltonian} {can increase the embedding and physical-qubit overhead on a quantum annealer.}
The proposed method below is designed to keep the advantage of UP without slack variables while removing the static squared residual from {the Hamiltonian used for sampling}.

\section{\label{sec:upom}Proposed Method: Unbalanced Penalization Ohzeki Method}

\subsection{\label{subsec:upom_formulation}Formulation}

We now combine UP with the Ohzeki method.
For clarity we present the case with a single constraint in Eq.~\eqref{eq:ineq_constraint}.
The extension to multiple constraints is obtained by assigning one auxiliary variable to each inequality.
Original UP contains the residual square $h(\bm{x})^2=\left(B-F(\bm{x})\right)^2$ in
Eq.~\eqref{eq:up_penalty}.
The Ohzeki method replaces this quadratic residual term by a linear coupling between the residual and an auxiliary variable.
With the residual convention $h(\bm{x})=B-F(\bm{x})$, the Hamiltonian used for sampling in \upom, {with constant terms dropped,} can be written as
\begin{equation}
  \widetilde{H}_{\rm U}(\bm{x};v^{(t)})
  =
  f_0(\bm{x})
  -\lambda_1 h(\bm{x})
  +v^{(t)}h(\bm{x}) .
  \label{eq:upom_residual_hamiltonian}
\end{equation}
Thus, the original UP term $\lambda_2h(\bm{x})^2$ is not kept in the sampler.
It is replaced by the auxiliary linear term $v^{(t)}h(\bm{x})$, whose coefficient is updated iteratively.
Because $h(\bm{x})=B-F(\bm{x})$, constants independent of $\bm{x}$ can be omitted and the two linear residual coefficients can be absorbed into a single effective coefficient
$u^{(t)}=v^{(t)}-\lambda_1$.
The sampler therefore uses the equivalent Hamiltonian
\begin{equation}
  H_{\rm U}(\bm{x};u^{(t)})
  =
  f_0(\bm{x})-u^{(t)}F(\bm{x}),
  \label{eq:upom_hamiltonian}
\end{equation}
or equivalently
\begin{equation}
  Q_{\rm U}(\bm{x};u^{(t)})
  =
  \frac{1}{Z_{\rm U}(u^{(t)})}
  \exp\left[
    -\beta f_0(\bm{x})
    +\beta u^{(t)}F(\bm{x})
  \right].
  \label{eq:upom_distribution}
\end{equation}
The auxiliary variable is updated by
\begin{equation}
  u^{(t+1)}
  =
  u^{(t)}
  +\eta_t
  \left(
    B-\Braket{F(\bm{x})}_{Q_{\rm U}(\bm{x};u^{(t)})}
  \right).
  \label{eq:upom_update}
\end{equation}

  Equation~\eqref{eq:upom_update} has the same form as the Ohzeki method for equality constraints, but the target is now the boundary of the inequality.

% Equation~\eqref{eq:upom_update} has the same form as the Ohzeki method for equality constraints, but the target is now the boundary of the inequality.
% If the sampled average exceeds the capacity, the residual in Eq.~\eqref{eq:upom_update} becomes negative and $u^{(t)}$ decreases, suppressing large-$F$ configurations in the next sampling step.
% If the sampled average is below the capacity, $u^{(t)}$ increases and the sampler is allowed to approach the boundary.

% This update is particularly natural for knapsack problems, where feasible solutions with high value usually lie near the capacity boundary.
% The best feasible sample observed during the iterations is stored as the solver output.
% Thus, infeasible samples are permitted during the search, but the returned candidate always satisfies the original inequality whenever a feasible sample has been observed.

\upom\ differs from original UP in two ways.
  First, it avoids the difficult two-dimensional hyperparameter search over $\lambda_1$ and $\lambda_2$ required in original UP.
The linear UP coefficient and the auxiliary coefficient are combined into $u^{(t)}$, while the role of the squared coefficient is replaced by the Ohzeki method and its step size schedule.
  Thus, the {tuning} is reduced from selecting two static penalty coefficients to choosing, or learning, a single update schedule.
Second, the sampler Hamiltonian in Eq.~\eqref{eq:upom_hamiltonian} contains only a linear coupling to $F(\bm{x})$.
The squared residual term in Eq.~\eqref{eq:up_hamiltonian} is therefore absent from the sampling problem.
This preserves the slack-free advantage of UP and can reduce the logical-variable and embedding overhead on a quantum annealer by avoiding both slack variables and the
  squared residual term in the sampler Hamiltonian.
The same sampling statistics also give the derivative needed for DU.
{For notational simplicity, we abbreviate $Q_{\rm U}(\bm{x};u^{(t)})$ as $Q_{\rm U}$ in the following equations.}
{For a single inequality constraint, the derivative required for DU is given by}
\begin{equation}
  \frac{\partial}{\partial u}
  \Braket{F(\bm{x})}_{Q_{\rm U}}
  =
  \beta
  \operatorname{Var}_{Q_{\rm U}}\!\left[F(\bm{x})\right].
  \label{eq:upom_variance}
\end{equation}
For {multiple} inequalities $F_k(\bm{x})\le B_k$, the derivative becomes the covariance matrix, {which is given by}
\begin{equation}
  \frac{\partial}{\partial u_l}
  \Braket{F_k(\bm{x})}_{Q_{\rm U}}
  =
  \beta
  \operatorname{Cov}_{Q_{\rm U}}\!\left(F_k(\bm{x}),F_l(\bm{x})\right).
  \label{eq:upom_covariance}
\end{equation}

\subsection{\label{subsec:upom_experiment}Numerical results}

In this subsection, we numerically compare original UP with the proposed \upom.
We evaluate the method on random binary knapsack instances,
\begin{equation}
  \begin{aligned}
  \max_{\bm{x}\in\{0,1\}^{N}} \quad &
  \sum_{i=1}^{N} p_i x_i \\
  \textrm{s.t.}\quad &
  \sum_{i=1}^{N} w_i x_i \le W .
  \end{aligned}
  \label{eq:knapsack_max}
\end{equation}
In the sampler we use the equivalent minimization objective
\begin{equation}
  f_0(\bm{x})=-\sum_{i=1}^{N}p_i x_i,
  \quad
  F(\bm{x})=\sum_{i=1}^{N}w_i x_i .
  \label{eq:knapsack_min}
\end{equation}
  Here $p_i$ and $w_i$ denote the profit and weight of item $i$, respectively, and $W$ is the knapsack capacity.
  {In the numerical experiments, we set $N=200$ and $W=50$ and independently draw each weight and profit as $w_i\sim\mathcal{U}(0.1,10)$ and $p_i\sim\mathcal{U}(0.1,100)$, respectively, where $\mathcal{U}(a,b)$ denotes the uniform distribution on $[a,b]$.}
% The instance distribution is $N=200$, $W=50$, $w_i\sim\mathcal{U}(0.1,10)$, and $p_i\sim\mathcal{U}(0.1,100)$.
The sampled weights and profits are truncated to three decimal places.
For each instance, {a reference optimal solution and its objective value are} obtained by {exact} dynamic programming.
In this computation, the weights and profits are multiplied by $1000$ so that the truncated data are represented as integers.
All reported comparisons use $500$ random instances and inverse temperature $\beta=0.5$.
Although \upom\ is motivated by implementation on quantum annealers, as a first step, we compare all methods using MH sampling at a fixed inverse temperature to assess the performance of the proposed methods without the influence of annealing-specific settings. Both SQA and QA are affected by the annealing schedule, while SQA is additionally affected by the number of Trotter slices.  
{The use of MH in the evaluation is also motivated by our previous study on equality-constrained problems, where replacing MH with SQA or QA did not degrade the convergence of either the fixed-step Ohzeki method or DUOM with learned step sizes~\cite{cqtl}. Thus, MH provides a reasonable first-step benchmark for assessing the relative convergence behavior of the proposed methods. Evaluations using SQA and a quantum annealer are left for future work.}

We use two evaluation metrics.
The first is the mean squared error to a reference optimal bit string,
\begin{equation}
  \MSE^{(t)}
  =
  \frac{1}{N}
  \norm{\hat{\bm{x}}^{(t)}-\bm{x}^{\ast}}_2^2,
  \label{eq:mse_metric}
\end{equation}
where $\hat{\bm{x}}^{(t)}$ is the best candidate recorded up to the current computational budget and $\bm{x}^{\ast}$ is {the corresponding reference optimal solution}.
The second is the optimal instance rate, defined as the fraction of test instances for which the best feasible value has reached {the reference optimal value}.

\begin{figure}[t]
  \centering
  \includegraphics[width=\columnwidth]{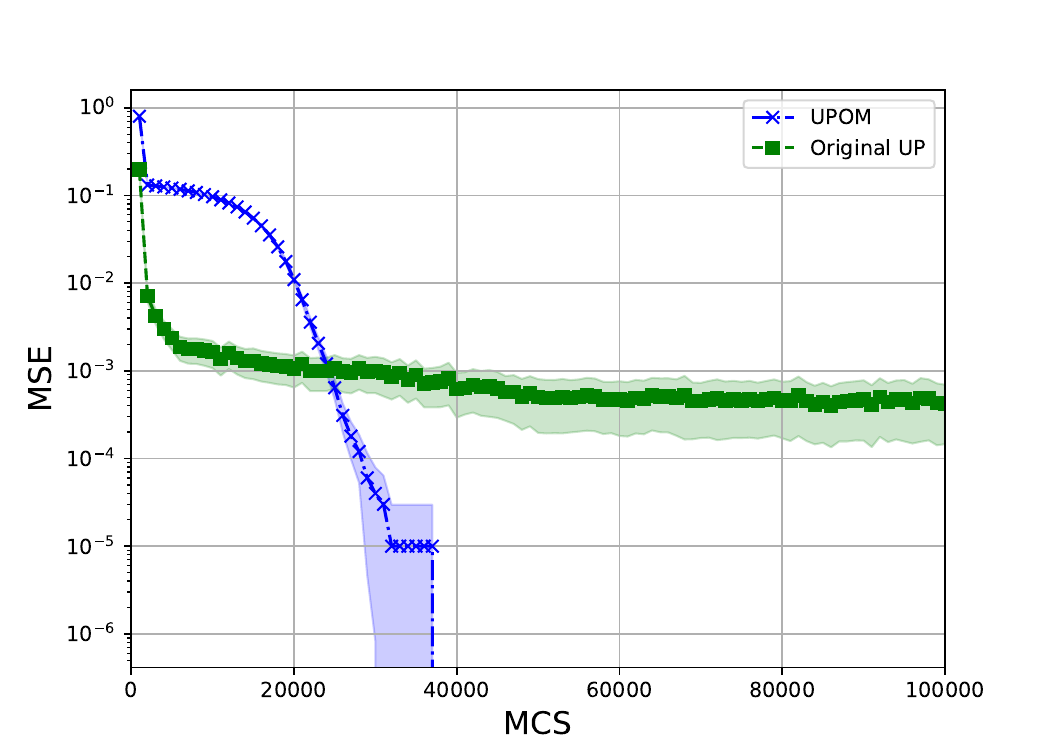}
  \caption{
    MSE comparison between original UP and \upom\ on random knapsack instances.}
  \label{fig:original_up_mse}
\end{figure}

\begin{figure}[t]
  \centering
  \includegraphics[width=\columnwidth]{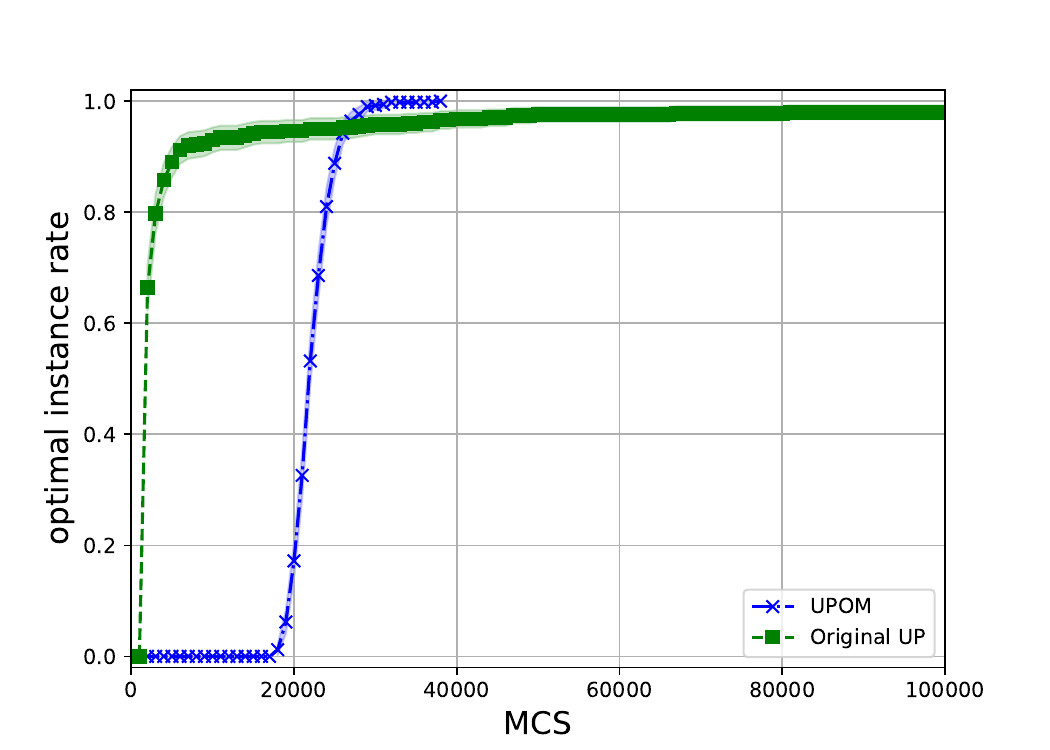}
  \caption{
    Optimal instance rate for original UP and \upom.}
  \label{fig:original_up_optimal}
\end{figure}

For original UP, we use Optuna~\cite{Optuna} to search the two hyperparameters in Eq.~\eqref{eq:up_hamiltonian}.
The hyperparameter search uses $\beta=0.5$, $5$ tuning instances, $50000$ Monte Carlo steps (MCS), $100$ trials, a recording interval of $1000$ MCS, and the final objective value as the optimization target.
  The values used in the final evaluation, rounded to four significant digits, are {$\lambda_1=29.41$ and $\lambda_2=3.129$}.
% \begin{equation}
%   \lambda_1=29.410387595926487,
%   \quad
%   \lambda_2=3.12927315020543 .
%   \label{eq:original_up_lambdas}
% \end{equation}
For \upom, we use a constant step size and initial auxiliary variable $u^{(0)}=1.0$.
To compare this with baselines without training, we conduct a grid search over fixed step sizes
\(\eta_t=5.0\times10^{-1}\), \(1.0\times10^{-1}\), \(5.0\times10^{-2}\), \(1.0\times10^{-2}\), \(5.0\times10^{-3}\), \(1.0\times10^{-3}\),  \(5.0\times10^{-4}\) and \(1.0\times10^{-4}\) for \(t=0,\ldots,T-1\).
We report the result with {the fixed step size that performed best}.
The selected value is $\eta_t=1.0\times10^{-1}$ for all $t$.
One \upom\ iteration uses $1000$ MCS.
The original UP and \upom\ curves in Fig.~\ref{fig:original_up_mse} and Fig.~\ref{fig:original_up_optimal} are therefore compared using the cumulative number of MCS on the horizontal axis.
{Unless otherwise stated, all error bars in this paper represent $95\%$ confidence intervals over $500$ random instances.}

Figure~\ref{fig:original_up_mse} shows that original UP does not reach zero MSE for all test instances within $100000$ MCS.
In contrast, \upom\ reaches zero MSE for all instances within $40000$ MCS, indicating that {it finds the reference optimal solution for every test instance}.
Figure~\ref{fig:original_up_optimal} shows the same tendency from the viewpoint of optimal instance rate.
Original UP increases rapidly during the first $10000$ MCS, but it does not reach $100\%$ because some instances remain unsolved.
\upom\ starts to increase sharply around $20000$ MCS and reaches $100\%$ optimal instance rate within $40000$ MCS.

This comparison highlights the first advantage of \upom.
Even after Optuna tuning, original UP still suffers from the difficulty of selecting two hyperparameters.
\upom\ replaces this static search over two parameters with an auxiliary variable update of the Ohzeki method.
\upom\ without training {requires selecting} only one fixed step size, and a simple grid search is sufficient to obtain good performance in this experiment.
{The Hamiltonian used for sampling} contains neither the squared residual term nor slack variables.
\upom\ is therefore a reformulation of UP that avoids the two-parameter penalty tuning required by original UP while preserving its slack-free design.

\section{\label{sec:duupom}Proposed Method: Deep Unfolded Unbalanced Penalization Ohzeki Method}

The numerical results in Sec.~\ref{subsec:upom_experiment} show that \upom\ gives a practical improvement over original UP.
We next apply the DUOM framework reviewed in Sec.~\ref{subsec:duom_background} to the UPOM update in Eq.~\eqref{eq:upom_update}.

\subsection{\label{subsec:duupom_application}{Application of Deep Unfolding to UPOM}}

{We propose \duupom\ by combining UPOM with DU.}
Its trainable parameters are the step sizes $\{\eta_t\}_{t=0}^{T-1}$ that depend on the layer in the UPOM auxiliary variable update.
The same backpropagation mechanism based on sampling described for DUOM is used, because the derivative of the UPOM expectation is given by the variance in Eq.~\eqref{eq:upom_variance}.
For the knapsack experiments, \duupom\ minimizes {the following expected penalized objective under the distribution after $T$ iterations:}
   \begin{equation}
    L_{\rm DU\text{-}UPOM}
    =
    \Braket{
    -\sum_{i=1}^{N}p_i x_i
    +
    \rho
    \left[
      \sum_{i=1}^{N}w_i x_i-W
    \right]_+^2
    }_{Q_{\rm U}}.
    \nonumber
  \end{equation}
Here $[a]_+=\max\{0,a\}$ denotes the positive part, and we set $\rho=6$.

\subsection{\label{subsec:reference_baselines}Reference baselines}

We compare \duupom\ with three reference methods without training.
The first is \upom\ with a fixed step size, which uses Eq.~\eqref{eq:upom_update} with a constant step size selected from the grid described in Sec.~\ref{subsec:upom_experiment}.
The other two methods apply the Ohzeki method to inequality constraints in ways different from \upom.
One is {the method proposed in} Ref.~\cite{Takabayashi2025}, {which combines Lagrangian relaxation with a projected subgradient update. We call this method the Lagrangian Ohzeki method (\lom) in this paper}.
{For a COP with one inequality constraint}, \lom\ {uses} a Lagrangian Hamiltonian of the form
\begin{equation}
  H_{\rm L}(\bm{x};\tau^{(t)})
  =
  f_0(\bm{x})+\tau^{(t)}F(\bm{x}),
  \quad
  \tau^{(t)}\ge0,
  \label{eq:lom_hamiltonian}
\end{equation}
{The corresponding probability distribution is}
\begin{equation}
  {
  Q_{\rm L}(\bm{x};\tau^{(t)})
  =
  \frac{1}{Z_{\rm L}(\tau^{(t)})}
  \exp\left[
    -\beta H_{\rm L}(\bm{x};\tau^{(t)})
  \right].
  }
\end{equation}
\lom\ {then} updates the multiplier by a projected step,
\begin{equation}
  \tau^{(t+1)}
  =
  \max\left\{
    0,\,
    \tau^{(t)}
    +\eta_t
    \left(
      \Braket{F(\bm{x})}_{{Q_{\rm L}(\bm{x};\tau^{(t)})}}-B
    \right)
  \right\}.
  \label{eq:lom_update}
\end{equation}
The other is the Ohzeki method with slack variables, which combines the standard encoding using slack variables~\cite{KomiyamaSuzuki2024} with the Ohzeki method for the resulting equality constraint~\cite{YuNabil2021}.
It introduces binary variables encoding a slack $S$ and applies the Ohzeki method to the equality $F(\bm{x})+S=B$.
Both \lom\ and the method using slack variables are important references, but they are not the main target of this paper.
The central comparison is whether DU improves the convergence speed of \upom.

\subsection{\label{subsec:duupom_experiment}Numerical results}

The \duupom\ experiment uses the same random knapsack {setting} as Sec.~\ref{subsec:upom_experiment}.
All methods are evaluated on $500$ instances with $\beta=0.5$.
For the baselines without training, \upom, \lom, and the Ohzeki method with slack variables, we use the same grid search over fixed step sizes described in Sec.~\ref{subsec:upom_experiment}.
For all three baselines, {$\eta_t=1.0\times10^{-1}$ performed best}, and the reported curves use this value.

To implement unsupervised learning for \duupom, we generated datasets consisting of {minibatches} of random knapsack instances with $N=200$ items and capacity $W=50.0$.
\duupom\ was implemented using PyTorch 2.0.0~\cite{pytorch} and {trained using minibatches}.
In each parameter update, we used {minibatches} of size $10$.
The parameters $\{\eta_t\}_{t=0}^{T-1}$ were optimized using the Adam optimizer~\cite{Adam} to minimize the sampled objective value.
The inverse temperature was set to $\beta=0.5$.
The initial learning rate was set to $8.0\times10^{-4}$ and was decayed by a factor of $0.95$ after each generation using a StepLR scheduler.

In the training, incremental training~\cite{TISTA,Incremental1} was employed to prevent gradient vanishing.
Here, a generation denotes one stage of incremental training.
In incremental training, the step-size parameters $\{\eta_t\}_{t=0}^{T-1}$ are learned sequentially by gradually increasing the number of iterations in the unfolded model.
Trainable parameters are added step by step as the number of unfolded iterations increases.
In generation $\ell$, the model with $\ell+1$ UPOM iterations is trained to optimize the parameter set $\{\eta_t\}_{t=0}^{\ell}$.
The training consisted of $50$ generations, and each generation contained $10$ parameter updates.
Each training sampler call used $100$ MCMC samples.
The initial parameter values were set to $\eta_t=1.0\times10^{-1}$ for all $t$.

\begin{figure}[t]
  \centering
  \includegraphics[width=\columnwidth]{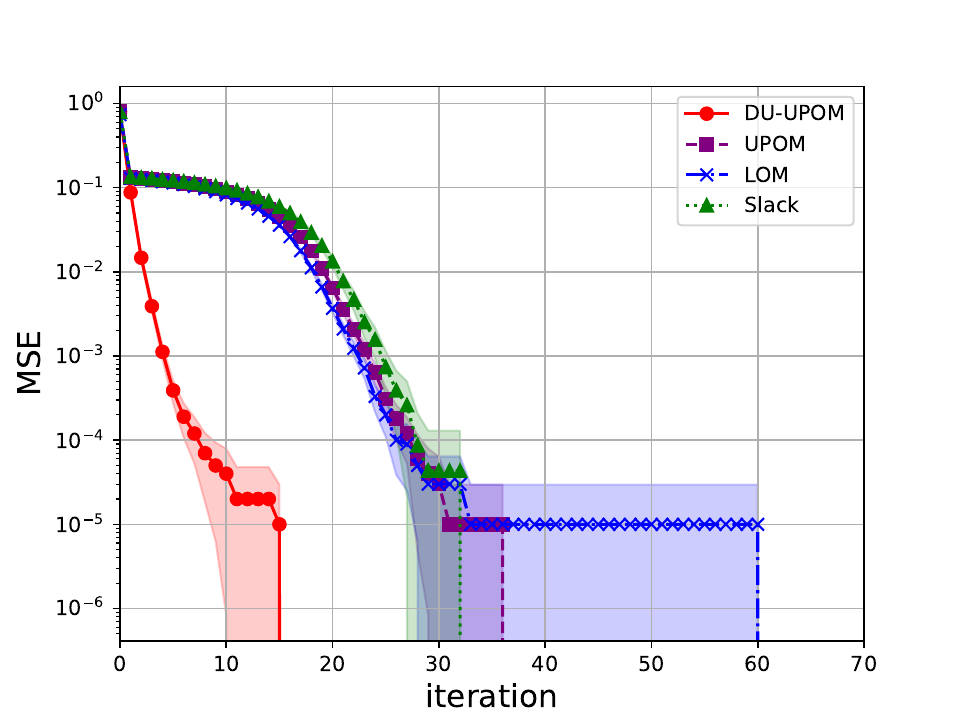}
  \caption{
    MSE comparison among \duupom, \upom\ with a fixed step size, \lom, and the Ohzeki method with slack variables.}
  \label{fig:duupom_mse}
\end{figure}

\begin{figure}[t]
  \centering
  \includegraphics[width=\columnwidth]{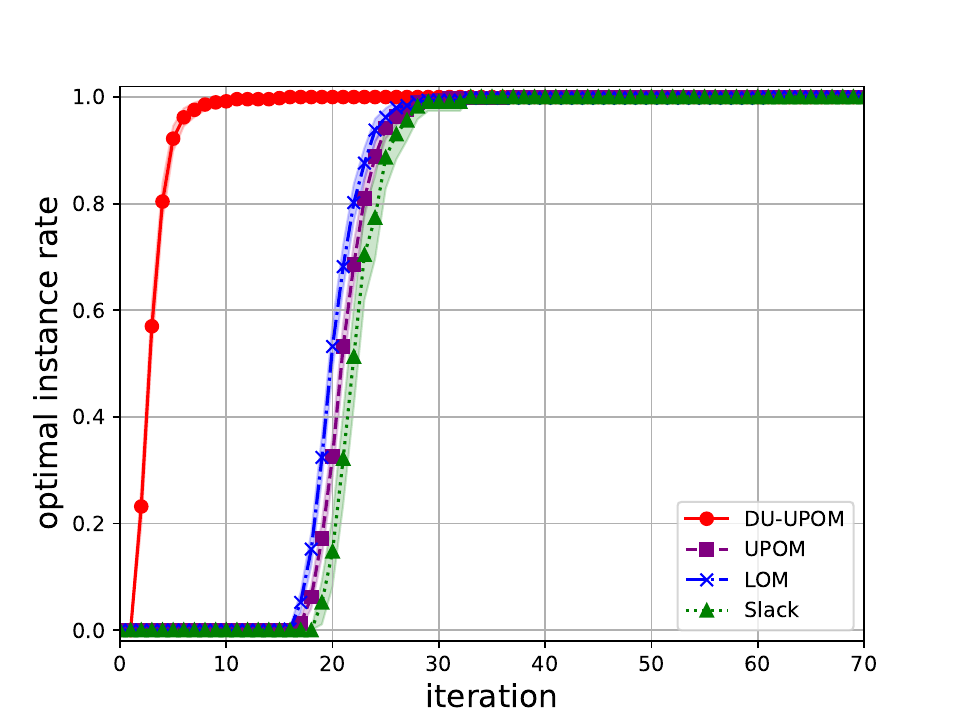}
  \caption{
    Optimal instance rate for \duupom, \upom\ with a fixed step size, \lom, and the Ohzeki method with slack variables.}
  \label{fig:duupom_optimal}
\end{figure}

Figure~\ref{fig:duupom_mse} shows that \duupom\ reaches zero MSE for all instances within $15$ iterations.
In contrast, the baselines without training, \upom\ with a fixed step size, \lom, and the Ohzeki method with slack variables, show similar convergence trends and require approximately $30$ iterations before almost all instances reach zero MSE.
We also observe that some instances remain trapped in local solutions for these baselines with fixed step sizes.
Figure~\ref{fig:duupom_optimal} shows the same tendency in terms of optimal instance rate.
\duupom\ reaches the optimal solutions much earlier, whereas the optimal instance rates of \upom\ with a fixed step size, \lom, and the Ohzeki method with slack variables increase more slowly and are affected by the instances that do not escape local solutions.

  These results show that \duupom\ improves the convergence speed of \upom\ by learning the step-size schedule {for} the auxiliary-variable update.
  Since each outer iteration requires MCMC sampling, reducing the number of iterations also reduces the number of sampler calls needed to obtain an optimal or high-quality feasible solution.

\begin{figure}[t]
  \centering
  \includegraphics[width=\columnwidth]{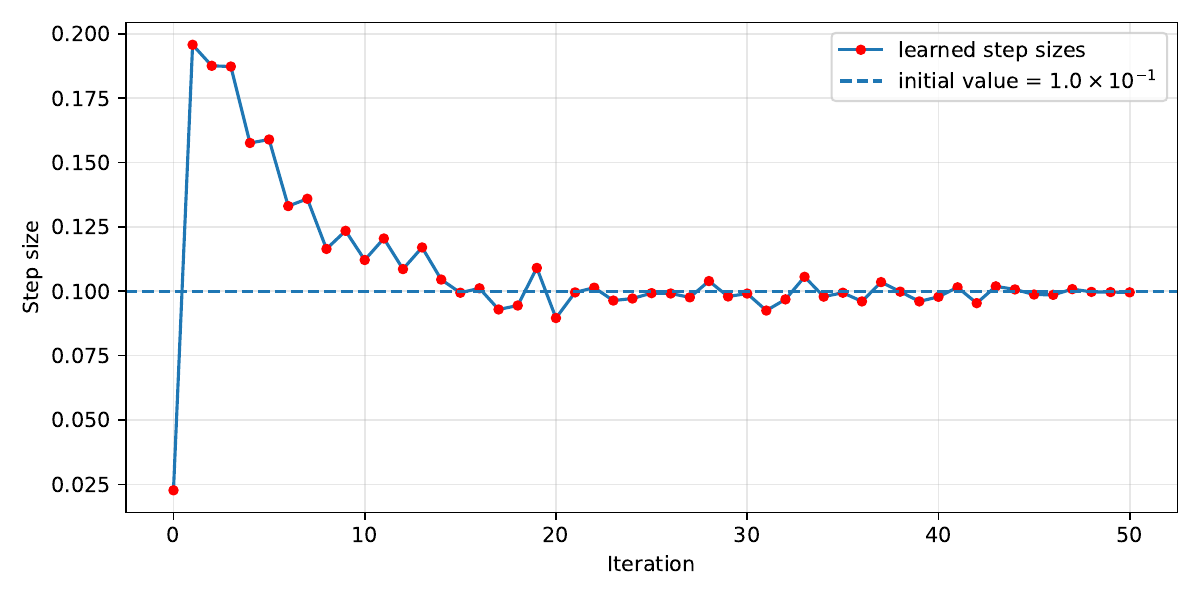}
  \caption{Step size schedule learned by \duupom. The learned values vary strongly with the iteration index, producing the nonuniform update schedule used in the evaluation.}
  \label{fig:step_sizes}
\end{figure}

  Figure~\ref{fig:step_sizes} shows the step-size schedule learned by \duupom.
  The learned values are clearly nonuniform across iterations, as often observed in deep-unfolded algorithms~\cite{Chebyshev,TISTA}.
   This indicates that \duupom\ does not merely replace the fixed step size by another constant value.
  Instead, it adapts the update strength at each iteration.

  These results show that DU can accelerate the UPOM framework by automatically tuning the step-size schedule.
  Together with the results in Sec.~\ref{subsec:upom_experiment}, this indicates that the proposed approach preserves the slack-free advantage of UP, removes the squared residual term from {the Hamiltonian used for sampling},
  and avoids the difficult two-parameter tuning of original UP.
  The comparisons with fixed-step \upom, \lom, and the slack-variable method confirm that the learned schedule improves convergence while retaining the motivation of reducing logical-variable and embedding overhead.

\par\bigskip
\section{\label{sec:conclusion}Conclusion}

We proposed \upom, a reformulation of UP using the Ohzeki method for COPs with inequality constraints.
Original UP is attractive because it avoids slack variables, but it requires two penalty hyperparameters whose joint tuning is difficult and includes a squared residual term.
Compared with original UP, \upom\ reduces the tuning task from two hyperparameters to one step size schedule.
It also removes the squared residual term from {the Hamiltonian used for sampling}.
Since this term {introduces} additional couplings that must be embedded on a quantum annealer, \upom\ can reduce the embedding and qubit overhead while keeping the variable set of UP without slack variables.

We also proposed \duupom, a deep-unfolded extension of \upom\ that learns the step-size schedule.
The derivative needed for training is expressed by variances and covariances under the sampled distribution, so the method can be trained even though the MCMC sampler itself is nondifferentiable.
Experiments on $500$ random knapsack instances showed that \upom\ reaches all optimal solutions faster than original UP after hyperparameter search, and that \duupom\ {achieves a high} optimal instance rate much faster than \upom\ with a fixed step size and the reference baselines using \lom\ and slack variables.
These results show that combining UP with the Ohzeki method improves the practical usability of UP.
DU further turns this formulation into a trainable solver by learning the step size schedule from problem instances.

Future work will focus on three directions.
First, we will apply \upom\ and \duupom\ to more practical COPs with inequality constraints beyond the random knapsack instances studied here.
Second, we will implement and evaluate {the proposed UPOM Hamiltonian} without slack variables on a quantum annealer, where the absence of slack variables and squared residual interactions is expected to be especially useful.
Third, we will analyze the convergence behavior and the learned step size schedules in more detail to clarify why DU accelerates the UPOM.

  \par\bigskip
\begin{acknowledgments}
This study was partially supported by JSPS KAKENHI with Grants No.~25K21297 {and No.~26K14992}.
\end{acknowledgments}

\end{document}